\documentclass[reprint,aps,prb,amsmath,amssymb,twocolumn,showkeys, superscriptaddress,floatfix,showkeys]{revtex4-2}
\usepackage{graphicx}
\usepackage{dcolumn}
\usepackage{bm}
\usepackage{bbm} 
\usepackage[colorlinks = true, allcolors = blue]{hyperref}
\usepackage{verbatim}
\usepackage{soul}
\usepackage{float}
\usepackage{xcolor}
\usepackage{multirow}
\usepackage{tikz}
\usetikzlibrary{shapes.geometric, arrows, shadows,positioning}

\newcommand{\dd}{\mathrm{d}}

\newcommand{\Det}[1]{\left\vert #1\right\vert}

\begin{document}
\title{Giant anisotropy and Casimir phenomena: the case of carbon nanotube metasurfaces}

\author{Pablo Rodriguez-Lopez}
\email{pablo.ropez@urjc.es}
\affiliation{{\'A}rea de Electromagnetismo and Grupo Interdisciplinar de Sistemas Complejos (GISC), Universidad Rey Juan Carlos, 28933, M{\'o}stoles, Madrid, Spain}
\affiliation{Laboratoire Charles Coulomb (L2C), UMR 5221 CNRS-University of Montpellier, F-34095 Montpellier, France}

\author{Dai-Nam Le}
\affiliation{Department of Physics, University of South Florida, Tampa, Florida 33620, USA}

\author{Igor V. Bondarev}
\affiliation{Department of Mathematics and Physics, North Carolina Central University, Durham, NC 27707, USA}

\author{Mauro Antezza}
\affiliation{Laboratoire Charles Coulomb (L2C), UMR 5221 CNRS-University of Montpellier, F-34095 Montpellier, France}
\affiliation{Institut Universitaire de France, 1 rue Descartes, F-75231 Paris Cedex 05, France}

\author{Lilia M. Woods} 
\email{lmwoods@usf.edu}
\thanks{Author to whom all correspondence should be addressed}
\affiliation{Department of Physics, University of South Florida, Tampa, Florida 33620, USA}

\date{\today}

\begin{abstract}

The Casimir interaction and torque are related phenomena originating from the exchange of electromagnetic excitations between objects. While the Casimir force exists between any types of objects, the materials or geometrical anisotropy drives the emergence of the Casimir torque. Here both phenomena are studied theoretically between dielectric films with immersed parallel single wall carbon nanotubes in the dilute limit with their chirality and collective electronic and optical response properties taken into account. It is found  that the Casimir interaction is dominated by thermal fluctuations at sub-micron separations, while the torque is primarily determined by quantum mechanical effects. This peculiar quantum vs. thermal separation is attributed to the strong influence of reduced dimensionality and inherent anisotropy of the materials. Our study suggests that nanostructured anisotropic materials can serve as novel platforms to uncover new functionalities in ubiquitous Casimir phenomena.
 
\end{abstract}
\maketitle

\section{Introduction}
The discovery of layered materials has elevated the importance of van der Waals (vdW) interactions as they are responsible for keeping their inert components together \cite{Gjerding_2021,Geng2018}. Advanced computational schemes have been implemented in state of the art density functional theory packages to take into account the vdW energy when simulating various materials properties \cite{Maurer2019,Kresse1996b,Giannozzi2017}. The materials aspects of the Casimir force, a retarded vdW interaction, has also generated significant interest \cite{Woods2016}. This ubiquitous interaction governs not only the performance of micro and nanomachines, but it also probes fundamental properties stemming from Dirac and topologically nontrivial physics \cite{Gong2021}. 

Another aspect of Casimir phenomena is the ability to generate Casimir torque when optically anisotropic materials are involved \cite{Broer2021,Antezza2020,Lu2016}. Indeed, the misalignment of the inequivalent optical axis of two bodies results in their relative rotation when electromagnetic (EM) fluctuations are exchanged. This type of motion has been recently demonstrated in the laboratory in birefringent liquid crystal systems \cite{Somers2018} giving further impetus of enhancing our possibilities to study basic physics via EM interactions. The main ingredient for a large Casimir torque is the materials strong anisotropy ensuring a substantial effect at various separations and temperature ranges. In this context, quasi-one dimensional structures, in which optical and geometrical anisotropies are combined, provide excellent conditions for angular dependence of the force and much enhanced Casimir torque \cite{Wang2021,Somers2018}.

Theoretical studies of the Casimir energy and torque rely on the Lifshitz formalism, where the relative orientation of the optical axis of materials separated by a distance $D$ is taken into account in the EM boundary conditions. Typically, the torque decays roughly as $D^{-3}$ and it has the characteristic $\sin (2\varphi)$ behavior. Also, its sign and magnitude depend on the optical properties of the materials \cite{Kenneth2001,Munday2005,Munday2006,Broer2023,Rodriguez-Lopez_2023}. Additionally, the angular dependence of the vdW and Casimir force has been explored in cylindrical quasi-one dimensional structures, for which optical levitation of a nanorod above a birefringent crystal has been proposed \cite{Rajter2007,Xu2017}.

In order to further exploit rotations generated by Casimir torque for manipulating micro and nano-machines, more studies are necessary to identify materials with strong anisotropy. In this regard, carbon nanotubes and metasurfaces containing single-wall carbon nanotubes (SWCNs) are quite suitable due to their quasi-one dimensionality. Such materials offer new applications in quantum electron transport, electron energy-loss spectroscopy, and mechanical reinforcement \cite{Hage2017,Tsukagoshi1999,Wu2021}. Ultrathin films composed of periodically arranged nanotubes have recently emerged as transdimensional materials with extraordinary optoelectronic properties \cite{He2016,Ho2018,Roberts2019,Zhu2023}. SWCN films can also support plasmon, exciton, and phonon-polariton eigenmodes. This brings novel aspects in light-matter interactions with nanotube chiralities, different mixtures, and dielectric background material as effective "knobs" of tunability \cite{Adhikari2021,Bondarev2021}. 

\begin{figure}[htbp]
    \centering
    \includegraphics[width = \columnwidth]{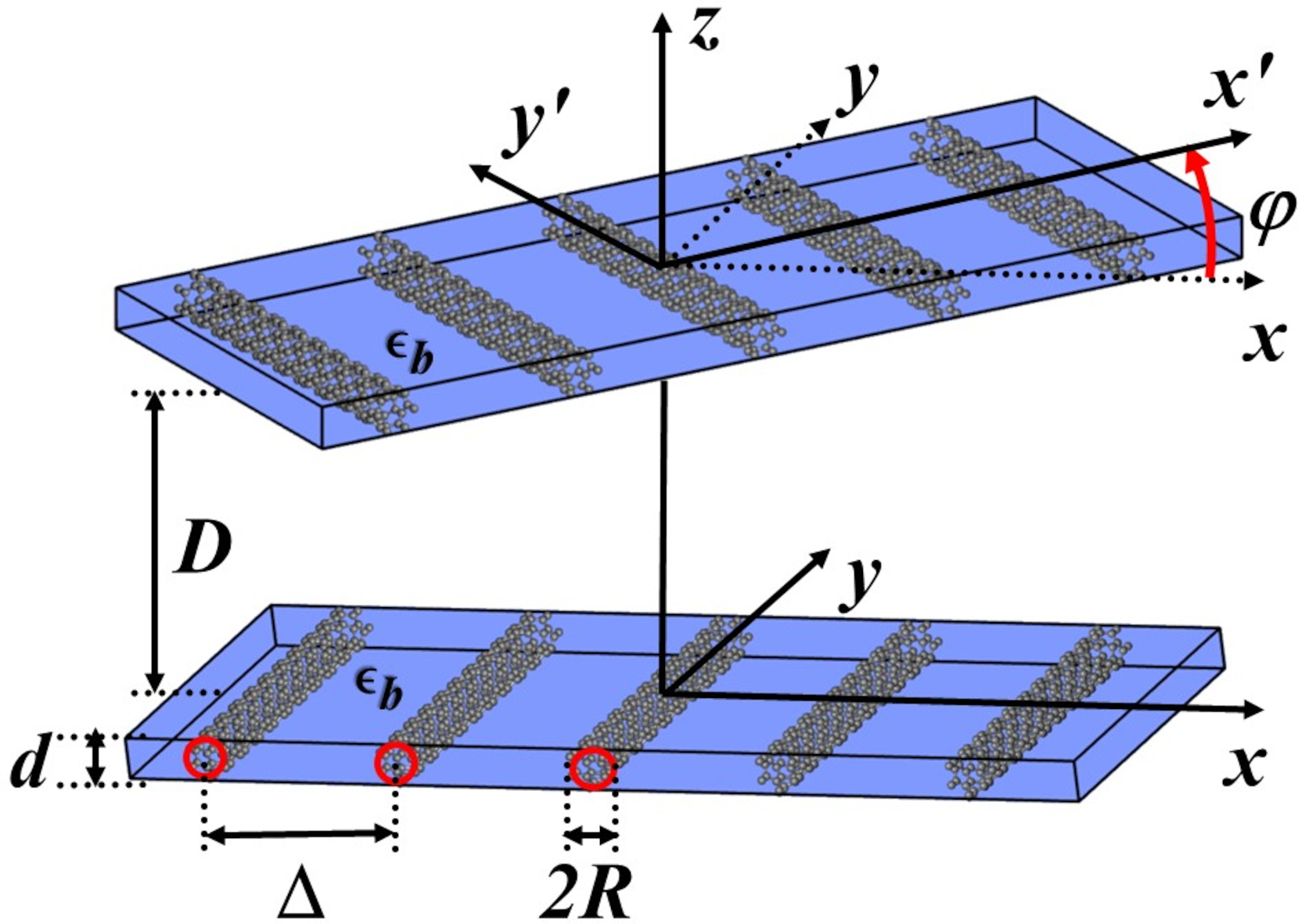}
    \caption{\label{fig:1} Two identical ultrathin SWCN films separated by a distance $D$ along the $z$-axis. The nanotube radius is $R$ and the intertube separation is $\Delta$. The SWCNs are imbedded in a solid dielectric layer with thickness $d \sim 2R \ll D$ and effective background dielectric constant $\epsilon_b$. Such films can be immersed in a liquid dielectric surroundings with a dielectric constant $\epsilon_s$ or be free-standing in air. The relative orientation of the nanotube axis from both structures is denoted by $\varphi$.}
\end{figure}

While much work has been devoted to properties of individual carbon nanotube films, light-matter interactions {\it between} such films has not been explored yet. Understanding how fluctuation-induced interactions occur, the factors that control their strength and characteristic behavior are questions of fundamental importance. In this paper, we investigate Casimir phenomena between two identical nanotube films using the Lifshitz formalism. By taking into account the nanotube optical response properties and the dielectric environment we show how the interplay between anisotropy, quantum mechanical and thermal fluctuations affect the ubiquitous Casimir force. The Casimir torque, a direct consequence of the anisotropy from the nanotube quasi-one dimensionality, is also investigated. Due to the reduced dimensionality and  properties of the  film, strong thermal fluctuations are found to play a dominant role in the interaction energy, while quantum mechanical effects are much more pronounced in the Casimir torque in the studied nm-$\mu$m separation limit.

\section{Properties of carbon nanotube thin films in the dilute regime}

We consider an ultrathin single wall carbon nano\-tube (SWCN) film composed of identical parallel aligned SWCNs embedded in a solid dielectric layer with thickness $d$ and effective background dielectric permittivity $\epsilon_b$ being a real constant. Such a film can be immersed in liquid dielectric surroundings with a dielectric constant $\epsilon_s$, or it can be free-standing in air. The SWCN array, schematically shown in Fig.~\ref{fig:1}, is aligned along the $y$ axis, and adjacent nano\-tubes are separated by a distance $\Delta$ in the $x$-direction (bottom material). The basic properties of each SWCN are captured by their chirality index $(n,m)$, which also determines the SWCN radius $R=\frac{\sqrt{3}b}{2\pi}\sqrt{m^2+nm+n^2}$ ($b=1.42$ $\AA\,$ is the C--C interatomic distance)~\cite{Saito1992}.

Since the SWCN radius ($R$ is in the nm range) is much smaller than its length (typically in the $\mu$m range), the optical response along the $y$-direction is essentially dominated by the collective longitudinal response of the array, while the response along the $x$-axis is entirely due to the dielectric medium. This in-plane anisotropic SWCN film can be treated as a quasi-2D system, where the vertical confinement due to the film thickness can be taken into account via an effective model using the Keldysh-Rytova potential~\cite{Keldysh1980}, as done in recent works for closely packed SWCN films~\cite{Adhikari2021,Bondarev2021}. Casimir interactions in densely packed SWCN film systems were recently studied in Ref. \cite{Bondarev2023}. The focus of our studies here is on the 'dilute SWCN film' regime, however, defined as $\Delta-2R\!\gg\!\epsilon_b d/(2\epsilon_s)$ with $d\!\sim\!2R$. In this case, the intertube electrostatic coupling is given by a $d$-independent 2D Coulomb interaction potential with a screening constant $\epsilon_s$ of the dielectric surrounding of the film~\cite{Keldysh1980}. As a consequence, the low-frequency (quasi-static) response of the SWCN array in the $y$-direction is predominantly due to individual SWCNs~\cite{Nakanishi2009}, while its higher-frequency optical (dynamical) response comes from the collective inter-tube exciton energy exchange due to induced dipole-dipole interactions~\cite{Bondarev2021}. The former can be represented by the properly normalized intraband surface conductivity of the individual constituent SWCN found in Ref.~\cite{Nakanishi2009}. The latter can be described by the dynamical surface conductivity of the SWCN array obtained from its collective excitonic response function reported recently in Ref.~\cite{Bondarev2021}.

Taking the above into account, the total surface conductivity of the SWCN array along the $y$-direction is (Gaussian units)
\begin{eqnarray}
\sigma_{yy}^{\text{array}}(k_y,\omega)=\frac{2\pi R}{\epsilon_s\Delta}\sigma_{yy}^{\text{intra}}(k_y,\omega)\hskip2.5cm\nonumber\\[-0.2cm]
\label{Sigma}\\[-0.2cm]
+\frac{\epsilon_b d}{2\pi}\frac{i\omega K(k_y)\sigma_{yy}^{\text{inter}}(k_y,\omega)}{i\omega+K(k_y)\sigma_{yy}^{\text{inter}}(k_y,\omega)}.\nonumber
\end{eqnarray}
The first term in the above expression is the intraband surface conductivity contribution coming from the single-tube intraband conductivity per unit surface,
\begin{eqnarray}
\sigma_{yy}^{\text{intra}}(k_y,\omega)=-\dfrac{2\alpha c v_{F}}{\pi^2R}\frac{i\omega-1/\tau}{(i\omega-1/\tau)^{2} + (v_{F}k_y)^{2}},
\end{eqnarray}
where $k_y$ is the absolute value of the electron quasi-momentum along the SWCN axis, $\alpha\!=\!e^2/\hbar c$, $v_F\!=\!c/300$ is the electron Fermi velocity in graphene, and $\tau$ is the phenomenological relaxation time parameter. This can be obtained by dividing its analogue per unit length of Ref.~\cite{Nakanishi2009} by $2\pi R$. The dimensionless prefactor $2\pi R/\Delta$ represents the surface fraction of the aligned SWCNs in the array.

\begin{figure}[htbp]
    \centering
    \includegraphics[width = 0.8 \columnwidth]{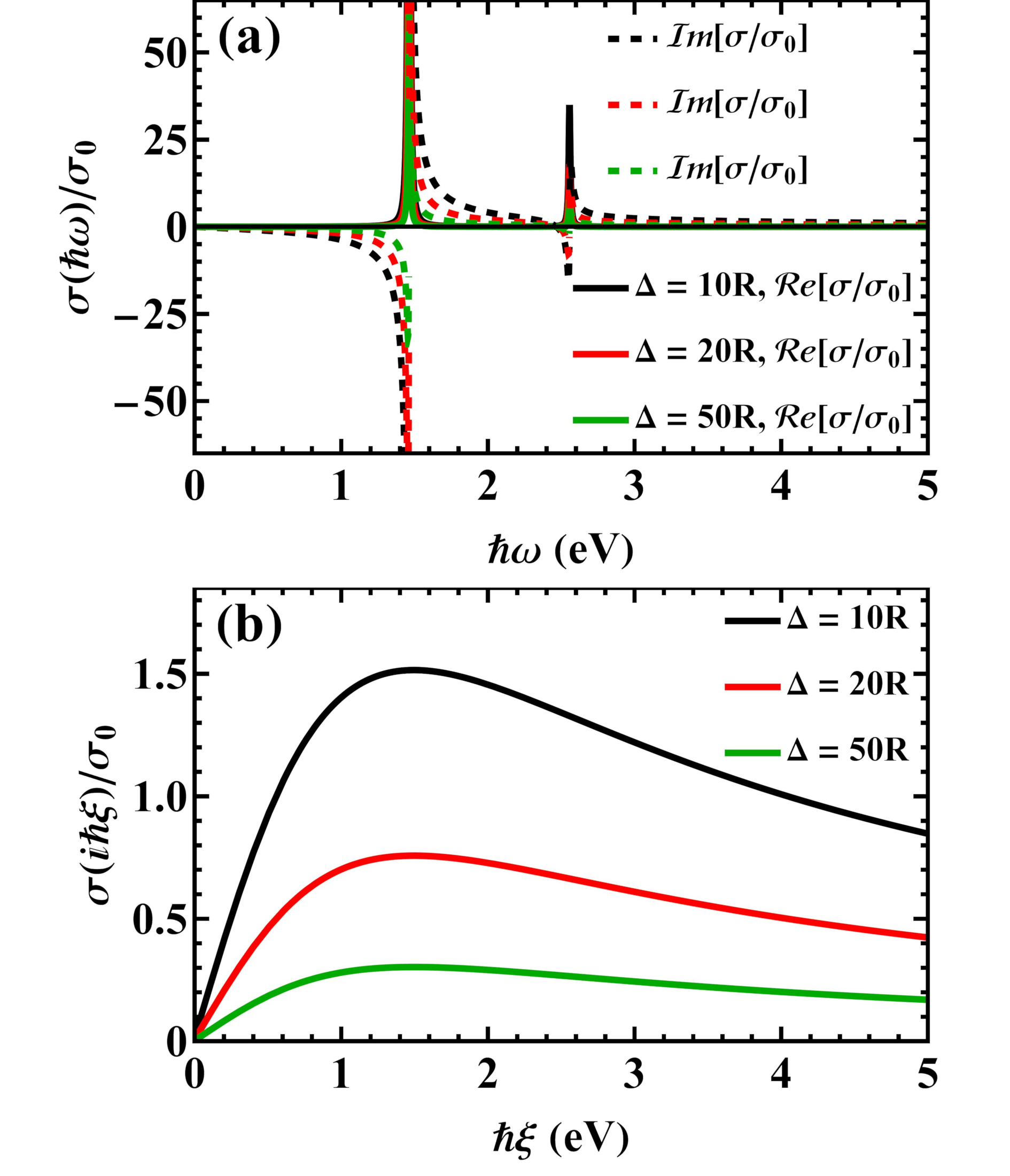}
    \caption{\label{fig:2} Scaled by  $\sigma_0 = \alpha c/4$ optical conductivity along the $y$-axis for a carbon film composed of (12,0) SWCNs at different separations $\Delta$ in (a) real frequency and (b) imaginary frequency domains. Here $k_{y} = 1/R$ and $\hbar/\tau=6.61$ meV corresponding to $\tau=100$ fs \cite{Perebeinos2005,Lazzeri2005}.}
\end{figure}

The second term in Eq.(\ref{Sigma}) is the interband surface conductivity contribution and it can be found  from its 3D equivalence $\sigma(\omega)=(-i\omega/4\pi))(\epsilon(\omega)-\epsilon)$ that relates the conductivity per unit volume to the dynamical response function of bulk isotropic material with static permittivity $\epsilon$ (see, e.g., Ref.~\cite{Kittel}). In this relationship, the right-hand side must be multiplied by $d$ to obtain conductivity per unit surface, the in-plane collective optical response $\epsilon(k_y,\omega)$ must be used for 
$\epsilon(\omega)$ and $\epsilon_b$ for $\epsilon$. The in-plane collective response function $\epsilon(k_y,\omega)$ in the alignment direction of the finite-thickness periodically aligned SWCN films was obtained in Ref.~\cite{Bondarev2021} using the many-particle Green's function formalism in the Matsubara formulation combined with the effective model based on the Keldysh-Rytova potential to include nonlocal plasmonic effects due to strong vertical confinement in ultrathin films systems~\cite{BondarevShalaev2017,Bondarev2019,Bondarev2020,Shah2022,BiehsBondarev2023,Salihoglu2023}. This yields the interband contribution to the SWCN array surface conductivity in Eq.(\ref{Sigma}), where $\sigma_{yy}^{\text{inter}}(k_y,\omega)$ is the interband conductivity of the individual SWCN (metallic or semiconducting) constituent, which can be found numerically using the Kubo formula and $({\bf k} \cdot {\bf p})$ method of the electronic band structure calculations~\cite{Ando2005}, and the function
\begin{equation}
K(k_y)=f_{CN}\frac{m^{*}\omega_{p}^{2}(k_y)d}{e^{2}N_{2D}R}
\label{Kq}
\end{equation}
with $f_{CN}\!=\!\pi R^2/(\Delta d)$ being the volume fraction of the SWCNs in the film. In essence, $K(k_y)$ captures the effect of collective oscillations of the surface electron density along the SWCN alignment direction with nonlocal plasma frequency
\begin{equation}
\omega_p(k_y)=\sqrt{\frac{4\pi e^2N_{\rm 2D}}{m^{\ast}\epsilon_b\,d}\frac{2k_yRI_0(k_yR)K_0(k_yR)}{1+2\epsilon_s/(\epsilon_b k_y d)}\,},
\label{omegap}
\end{equation}
where $m^\ast$ is the electron effective mass, $N_{\rm 2D}$~$(=\!N_{\rm 3D}d)$ is the surface electron density, $k_y$ is the absolute value of the quasi-momentum along the nanotube axis, and $I_0$ and $K_0$ are the zeroth-order modified cylindrical Bessel functions responsible for the correct normalization of the electron density distribution over cylindrical surfaces~\cite{Bondarev2019}.

The conductivity component $\sigma_{xx}^{\text{array}}$ along the $x$-direction can similarly be obtained from the 3D expression as discussed earlier. Given that the SWCN response is negligible in directions transverse to the nanotube axis, one finds, 
\begin{equation}
\sigma_{xx}^{\text{array}}(\omega)=-d\frac{i\omega}{4\pi}(\epsilon_b - \epsilon_s).
\label{Sigmaxx}
\end{equation}
The above expression corresponds to the in-plane collective optical response function along the $x$-direction of the finite-thickness periodically aligned SWCNs. 

With all of the above, the two-component in-plane surface conductivity tensor of the diluted quasi-2D SWCN array takes the diagonal form,
\begin{equation}
\label{Cond-Tensor}
\hat{\sigma}=
\begin{pmatrix}
\sigma_{xx}^{\text{array}}(\omega)  &   0 \\
0              &  \sigma_{yy}^{\text{array}}(k_y,\omega)
\end{pmatrix}
\end{equation}
with its individual components defined by Eqs.~(\ref{Sigma})--(\ref{omegap}).

In Fig. \ref{fig:2}, we show the calculated optical conductivity along the $y$-axis for a film composed of (12,0) SWCNs, taken as an example system. For the chosen values of $\Delta$ it is assumed that the dilute regime $\frac{\epsilon_s}{\epsilon_b}\Delta\gg 2R$ holds (see earlier discussions). The optical transitions in the $\hbar \omega >1$ eV region result from the interband transitions of the collective excitations as shown in the above equations. We find that although smaller $\Delta$ moves the transitions towards smaller frequencies, the shift is rather minor. The main role of the intertube separation is much more pronounced in the strength of the transitions, which can also be seen in the imaginary frequency domain. As shown in Fig. \ref{fig:2}b, the larger nanotube density per unit area results in a stronger overall response, which is important for the Casimir phenomena as discussed below.

\onecolumngrid

\begin{figure}[H]
    \centering
    \includegraphics[width = 0.9 \columnwidth]{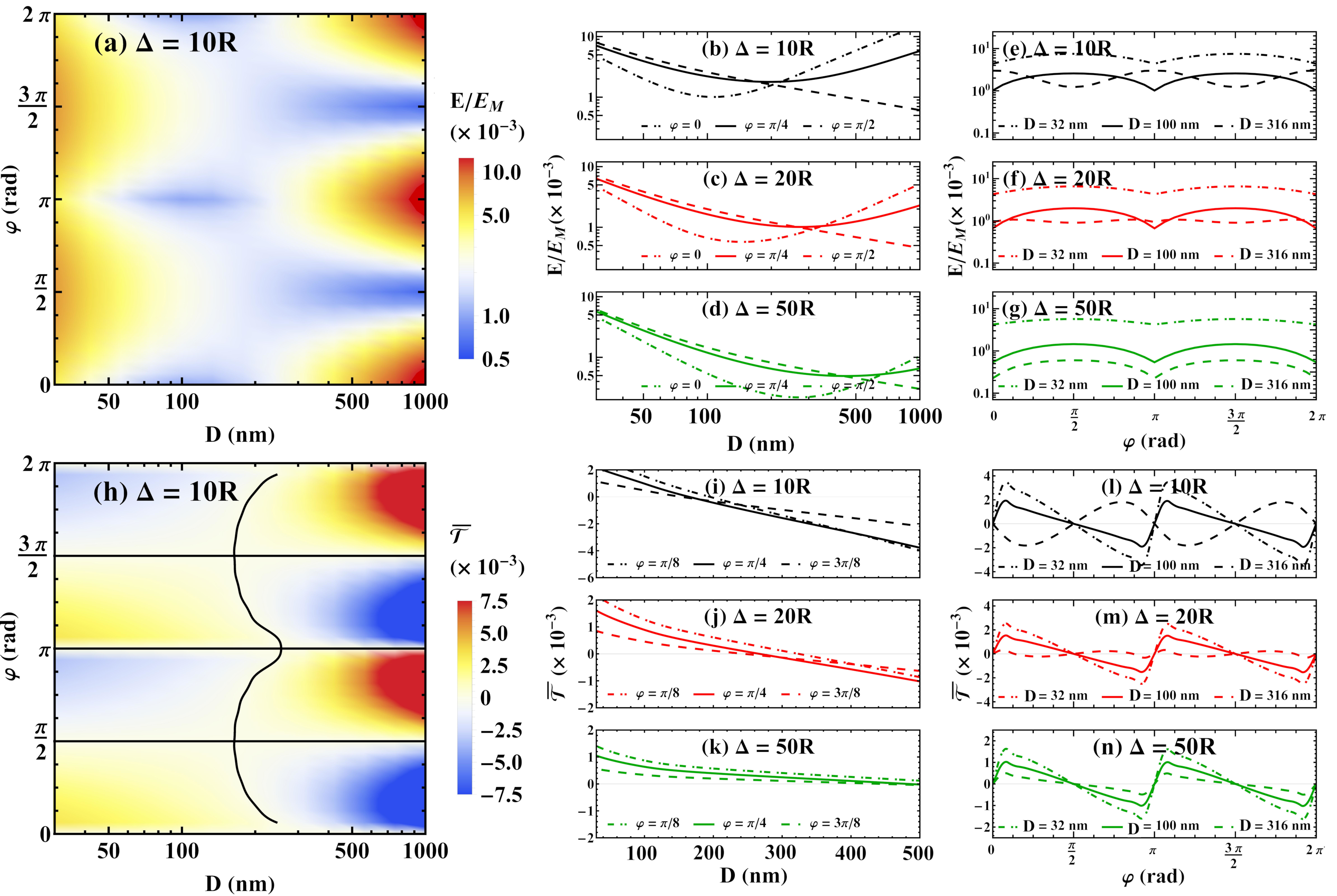}
    \caption{\label{fig:3} Density plots in $(\varphi, D)$ space of (a) Casimir energy $E$ normalized to $E_M=-\frac{\pi^2 \hbar c}{720 D^3}$ and (h) Casimir torque $\overline{\mathcal{T}}=\mathcal{T}/E_M$ for a SWCN film composed of (12,0) nanotubes with $\Delta=10R$. $E/E_M$ as a function of separation $D$ for (b) $\Delta=10R$, (c) $\Delta=20R$, (d) $\Delta=50R$ for optical axis relative orientation with $\varphi = \{0, \frac{\pi}{4}, \frac{\pi}{2}\}$. $E/E_M$ as a function of angle $\varphi$ for (e) $\Delta=10R$, (f) $\Delta=20R$, (g) $\Delta=50R$ for separation film between $D = \{32, 100, 316\}$ nm. $\overline{\mathcal{T}}$ as a function of separation $D$ for (b) $\Delta=10R$, (c) $\Delta=20R$, (d) $\Delta=50R$ for optical axis relative orientation with $\varphi = \{\frac{\pi}{8}, \frac{\pi}{4}, \frac{3\pi}{8}\}$. $\overline{\mathcal{T}}$ as a function of angle $\varphi$ for (e) $\Delta=10R$, (f) $\Delta=20R$, (g) $\Delta=50R$ for film separation $D = \{32, 100, 316\}$ nm. 
}
\end{figure}
\twocolumngrid

\section{Casimir effects at the quantum and thermal limits}

For the Casimir phenomena we consider two identical SWCN films (each with thickness $d=2R$) separated by a distance $D$ along the $z$-axis as displayed in Fig. \ref{fig:1} as each film is composed of the equally spaced (12,0) nanotubes. In addition to the Casimir energy per unit area $E$, a Casimir torque $\mathcal{T}= -\partial_{\varphi}E$ per unit area is also possible due to the optical anisotropy in this system set by the nanotube axis. Within the Lifshitz formalism the interaction energy and torque are found as
\begin{eqnarray}
\label{Energy}
& E = & k_B T \sideset{}{'}\sum_{n = 0}^{\infty}\int\frac{\dd^{2}\bm{k}_{\perp}}{\left(2\pi\right)^{2}} \nonumber\\
& & \times \ln\Det{ \mathbbm{1} - e^{-2 D \sqrt{ \kappa_{n}^{2} + \bm{k}_{\perp}^{2}}}\mathbb{R}_{0}\mathbb{R}_{\varphi} },\\  
\label{Torque}
&\mathcal{T} = & - k_{B}T \sideset{}{'}\sum_{n = 0}^{\infty} \int \frac{d^{2}\bm{k} _{\perp}}{(2\pi)^{2}}\times\nonumber\\
& & \times \text{tr}\left[\left( \mathbbm{1} e^{2 D \sqrt{ \kappa_{n}^{2} + \bm{k}_{\perp}^{2}}} - \mathbb{R}_0 \mathbb{R}_{\varphi} \right)^{-1} \mathbb{R}_0 \frac{d \mathbb{R}_{\varphi}}{d \varphi} \right].
\end{eqnarray}
The above expressions are obtained in imaginary Matsubara frequencies $\xi_{n} = c\kappa_{n}=2\pi n k_{B} T/\hbar$ and the prime in the summation corresponds to the $n=0$ term multiplied by $1/2$. The response properties of the materials, also taken in Matsubara frequencies, are captured in the Fresnel reflection matrices found from standard electromagnetic boundary conditions \cite{Rodriguez-Lopez_2023}

\begin{eqnarray}\label{Fresnel}
\mathbb{R}_{\varphi} = \frac{2\pi}{\delta_{\varphi}} \begin{pmatrix}
r^{xx}_{\varphi} & r^{xy}_{\varphi} \\
r^{yx}_{\varphi} & r^{yy}_{\varphi}
\end{pmatrix},
\end{eqnarray}
\begin{eqnarray}\label{Fresnel-2}
\left\{
\begin{array}{lll}
r^{xx}_{\varphi} &=& - 2\pi\left( \frac{\sigma_{yy}^{\varphi}}{c\lambda_n} + \frac{2\pi}{c}\Det{\sigma^{\varphi}}\right) \\
r^{xy}_{\varphi} &=& - \frac{2\pi}{c}\sigma_{yx}^{\varphi}  \\
r^{yx}_{\varphi} & = & \frac{2\pi}{c}\sigma_{xy}^{\varphi} \\
r^{yy}_{\varphi} & = & 2\pi\left( \lambda_n\frac{\sigma_{xx}^{\varphi}}{c} + \frac{2\pi}{c}\Det{\sigma^{\varphi}}\right)\\
\delta_{\varphi} & = & 1 + \frac{2\pi}{c}\left( \lambda_n\sigma_{xx}^{\varphi} + \frac{\sigma_{yy}^{\varphi}}{\lambda_n} \right) + \frac{4\pi^{2}}{c^{2}}\Det{\sigma^{\varphi}},
\end{array}
\right.,
\end{eqnarray}
where $\lambda_n = \sqrt{\bm{k}_{\perp}^2 c^2/\kappa_n^2 + 1}$ with $\bm{k}_{\perp}=(k_x, k_y)$ being the two-dimensional wave vector and $\Det{\sigma^{\varphi}} = \sigma_{xx}^{\varphi}\sigma_{yy}^{\varphi} - \sigma_{xy}^{\varphi}\sigma_{yx}^{\varphi}$. The conductivity tensor components of the rotated film can be obtained from $\sigma^{\varphi}=R_{\varphi, \hat{z}}^{-1}\sigma R_{\varphi, \hat{z}}$ where $\sigma$ is taken from Eq. \ref{Cond-Tensor} and $R_{\varphi,\hat{z}}=
\begin{pmatrix}
	\cos\varphi & -\sin \varphi \\
	\sin\varphi   &  \cos\varphi
\end{pmatrix}$ is the rotation matrix around the $z$-axis by an angle $\varphi$.  The Fresnel matrix $\mathbb{R}_0$ correspond to $\mathbb{R}_{\varphi = 0}$.  

In Eqs. \ref{Energy} and \ref{Torque} we distinguish between the quantum mechanical limit, in which the  summation over Matsubara frequencies $k_B T \sideset{}{'}\sum\limits_{n = 0}^{\infty}$ is replaced by the integral $\dfrac{\hbar c}{2 \pi}\int\limits_{0}^{\infty}d\kappa$, and the purely thermal limit, found from the $n=0$ Matsubara term. Both of these limits are examined in what follows.

The quantum mechanical Casimir energy $E_{qm}$ and torque $\mathcal{T}_{qm}$ are calculated numerically  using the Lifshitz expressions by taking the optical response model for the SWCN films, as discussed earlier. In Fig. \ref{fig:3}(a,h), we show $(\varphi, D)$ density maps with $\Delta=10R$ for both properties. Clear signatures of the nanotube anisotropy are noted in Fig. \ref{fig:3}(a). At larger separations, where thermal effects dominate, the interaction is strongest when the optical films are aligned ($\varphi = \{0, \pi, 2\pi\}$) and weakest when the optical axis of the fims are perpendicular to each other ($\varphi = \{\frac{\pi}{2}, \frac{3\pi}{2}\}$). At smaller $D$, however, this trend is reversed: the strongest coupling occurs for perpendicular optical axis, which was also recently found for the quantum Casimir force in densely packed SWCN films \cite{Bondarev2023}. To better understand the behavior of $E_{qm}$, we further show its dependence upon distance separation for several angles $\varphi$ and parameter $\Delta$. 

We find that as $D$ becomes larger and $\varphi \neq \{\frac{\pi}{2}, \frac{3\pi}{2} \}$, $E_{qm}$ approaches the limit of interacting metals with the characteristic $\frac{1}{D^3}$ scaling law, as shown in Fig. \ref{fig:3}(b-d). This is not surprising since the interaction in this range is dominated by the Drude-like response of the SWCN film. As $D$ becomes smaller, however, the energy experiences a transition to a $E_{qm}\sim\frac{1}{D^{4}}$ marking the onset dominance of the interband optical response in Eq. \ref{Sigma}. The parameter $\Delta$ and the particular distance at which the scaling transition happens have a positive correlation as shown in Fig. \ref{fig:3}(b-d). While this is the case for $\varphi = \{0, \frac{\pi}{4} \}$, for $\varphi=\frac{\pi}{2}$ the energy is markedly different. In this case, the $E_{qm}\sim\frac{1}{D^{4}}$ behavior (typical for the Casimir metal-dielectric interaction in 2D \cite{PhysRevLett.112.056804,NatComs_Lilia_2017}) is found in the entire distance range. Such orientation-dependent Casimir-Polder scaling laws have also been found in systems involving anisotropic particles \cite{Bimonte2015}, but to our knowledge  they have not been reported in two-dimensional anisotropic materials. The angular dependence of the Casimir energy is also shown in 
Fig. \ref{fig:3}(e, g). The oscillatory-like features are more pronounced for SWNT films with smaller $\Delta$ and smaller separations between the films. 

The density plot in Fig. \ref{fig:3}(h) shows that the Casimir torque in the quantum mechanical regime displays the characteristic $\sin (2\varphi)$ oscillations whose phase changes to $-\sin (2\varphi)$ at a certain distance. This can also be seen explictly in Fig. \ref{fig:3}(i-k). Our results indicate that the distance at which the torque experiences this phase change is closely related to the distance at which the energy changes its $D$ dependence (described earlier). The trend that the phase change occurs at smaller $D$ for smaller $\Delta$ is also observed in the Casimir torque. It appears that the interband-intraband terms and their relevance at different separation regimes are the main driving factor behind this effect. The $\sin (2\varphi)$ oscillations are explictly given in Fig. \ref{fig:3}(i-k), which correlate with the oscillatory features of the energy in  Fig. \ref{fig:3}(e-g). We also find that smaller $\Delta$ results in larger $E_{qm}$ and $\overline{\mathcal{T}}_{qm}$ as expected since denser nanotube arrays have stronger response properties. 

Casimir phenomena are also affected by temperature as thermal fluctuations may become prominent even at sub-micron sepations. This is the case especially for materials with reduced dimensions \cite{Gomez-Santos2009,Abbas2017,Khusnutdinov2018,Drosdoff2014,Drosdoff2016,Le2022}. To capture the role of temperature in the Casimir energy and torque of the films, we first begin by considering the $T$-dependence in the optical response properties. Many dielectric substances, such as teflon, polymers, or different types of glass  experience very weak temperature dependence in their dielectric properties \cite{Wongwilawan2013,Danewalia2016,Joseph2011}, thus here we assume that $\epsilon_b$ is $T$-independent. The intraband conductivity of the nanotube, however, is modified according to
\begin{eqnarray}
\sigma_{yy}^{\text{intra}}(k_y, \omega, T) &=& \sigma_{yy}^{\text{intra}}(k_y, \omega, 0) \nonumber\\
   &\times& \dfrac{k_{B}T}{\hbar v_{F}k_{y}}\ln\left( \dfrac{e^{(\mu - \hbar v_{F}k_{y})/k_{B}T}-1}{e^{\mu/k_{B}T}-1} \right),
\end{eqnarray}
where $\sigma_{yy}^{\text{intra}}(k_y, \omega, 0)$ is the intraband conductivity at $T=0$ (considered in the quantum limit calculations discussed earlier) and the chemical potential is taken to be $\mu=0.5$ eV \cite{Hung2015}. It is noted, however, that the dependence upon $\mu$ in the $\sigma_{yy}^{\text{intra}}(k_y, \omega, T)$ is relatively weak. 
The above expression can be obtained by using the Maldague formula \cite{Maldague1978}. On the other hand, the interband conductivity of the nanotube film is not significantly affected by temperature, as also shown in \cite{Bondarev2020,Vertchenko2019} where the combined effect of temperature and exciton-plasmon coupling in the individual nanotubes was considered.

The $T-$dependent Casimir properties can subsequently be calculated by using the Lifhsitz formalism with the explicit Matsubara summation in the energy and torque expressions from Eqs. \ref{Energy} and \ref{Torque}, respectively. The special $n=0$ term corresponds to the completely classical thermal regime and here it is also evaluated separately. We find that in this case one has $\mathbb{R}_{\varphi}(\kappa_n=0) = 
\begin{pmatrix}
	0 & 0 \\
	0  &  1
\end{pmatrix}$ regardless of the relative angle $\varphi$. As a result, the thermal Casimir energy is found as $E_{T}=-\frac{\zeta(3) k_{B}T}{16\pi D^2}$. It appears that in this classical thermal regime, the materials properties (including their anisotropy) are absent, and the Casimir energy is the same as the one for isotropic Drude metals. 

\begin{figure}[htbp]
    \centering
    \includegraphics[width = \columnwidth]{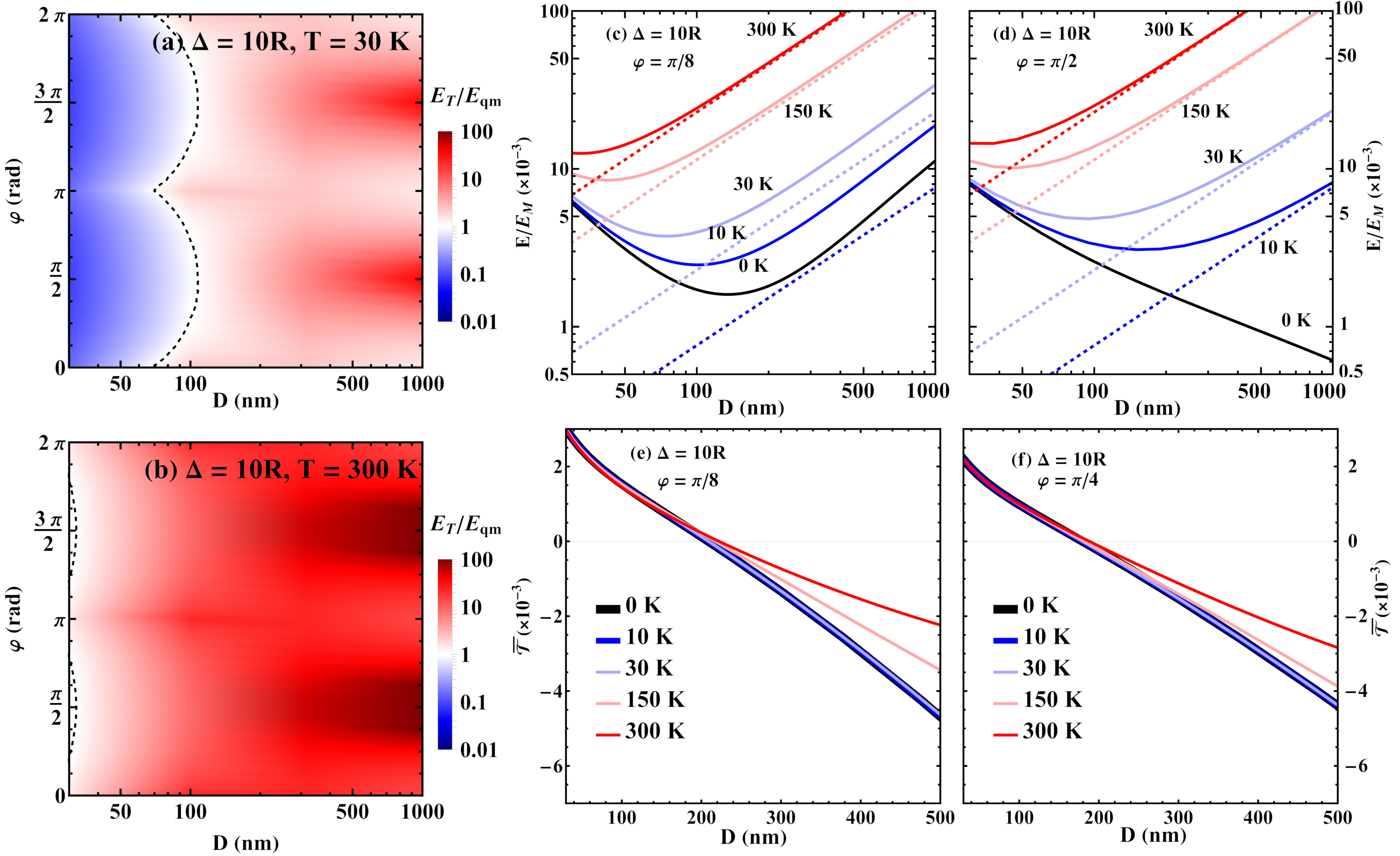}
    \caption{\label{fig:4} Density plots in ($\varphi, D$) space of the Casimir energy ratio $E_{T}/E_{qm}$ of the (12,0) CNT films with $\Delta=10R$ at: (a) $T=30$ K and (b) $T=300$ K. The Casimir energy ratio $E/E_{M}$ (obtained from Eq. \ref{Energy}) as a function of CNT film separation $D$ for (c) $\varphi=\pi/8$ and (d) $\varphi=\pi/2$ at different temperatures. The dashed lines correspond to the thermal limit from the $n=0$ Matsubara term. The Casimir torque ratio $\overline{\mathcal{T}}=\mathcal{T}/E_M$ (obtained from Eq. \ref{Torque}) as a function of separation $D$ for (e) $\varphi=\pi/8$ and (f) $\varphi=\pi/4$ at different temperatures.}
\end{figure}

In Fig. \ref{fig:4}(a,b), we give the $(\varphi,D)$ density plots of the ratio $E_{T}/E$ at $T=30$ and $T=300$ K, where $E$ is calculated from Eq. \ref{Energy}. These results show that temperature has a very strong effect on the Casimir energy. At lower $T$, quantum mechanical effects dominate the interaction energy at separation less than 100 nm, but for larger $D$ thermal effects become much more prominent. At higher temperatures, $E_{T}$ is much stronger even at $D<100$ nm. Another feature found here is that the strength of thermal effects compared to the quantum mechanical interaction depend on $\varphi$. For $\varphi = \{\frac{\pi}{2}, \frac{3\pi}{2} \}$, quantum mechanical effects are strong at smaller $D$ and thermal effects are strong at larger $D$ as can be seen even at $T=30$ K (\ref{fig:4}(a)). This transition shifts not only shifts towards smaller separations as $T=300$K, but the strength of thermal fluctuations become more prominent (\ref{fig:4}(b)). For $\varphi =\pi$, however, this trend does appear: thermal fluctuations have a diminished role at larger $D$, but they appear more prominent at intermediate separations. This type of non-uniform $\varphi$ dependence is associated with the phase change behavior in the energy entangled with the optical response of the film, as discussed earlier in this particular angular axis orientation.

\onecolumngrid

\twocolumngrid

These trends are shown in more details in Fig. \ref{fig:4}(c,d), where the Casimir energy is shown for $\varphi=\pi/8$ and $\pi/2$ but at several temperatures. One can see that at small $T$ the energy deviates significantly from $E_{T}$ in the displayed separation range for $\varphi=\pi/8$, however, for $\varphi=\pi/2$ the energy tracks $E_T$ at larger $D$. As $T$ increases, both orientations exhibit similar onset of thermal fluctuations as a function of distance separation. We further find that due the weak $T$-dependence of the optical response of the SWCN film, the Casimir energy at any temperature can be represented simply by adding the quantum mechanical $E_{qm}$ and the $n=0$ Matsubara term: $E\approx E_{qm}+E_T$. In fact, the results shown in Fig. \ref{fig:4}(c,d) obtained via Eq. \ref{Energy} completely overlap, which makes the Matsubara summation redundant. The distance for quantum-thermal effects transition in the Casimir energy is found by taking $E_{qm}=E_{T}$ and it is also displayed in Fig. \ref{fig:4}(a,b) (dashed black curves).

For the Casimir torque, we find that the classical $n=0$ term vanishes indicating the absense of classical thermal fluctuations. This unusual result is directly connected with the peculiar form of the Fresnel reflection matrix at zero Matsubara frequency $\mathbb{R}_{\varphi}(\kappa_n=0)$. For this special term, the anisotropy of the SWCN films is washed away meaning that the Casimir torque from purely thermal fluctuations is zero. In Fig. \ref{fig:4}(e,f), results are given for the Casimir torque as a function of separation for $\varphi = \{\frac{\pi}{8}, \frac{\pi}{4} \}$ obtained via Eq. \ref{Torque} for several temperatures. One finds that the thermal effect is rather different than in the case of the Casimir energy. For small distances, ($D<200$ nm in Fig. \ref{fig:4}(e,f)), the torque is completely determined by quantum fluctuations, while for larger distances, the torque can be approached by the $n=1$ Matsubara term, becoming exponentially suppresed ($\mathcal{T}\propto e^{-\frac{k_{B}T}{\hbar c}D}$) for distances $D>200$ nm and larger temperatures (\ref{fig:4}(e,f)).

\section{Conclusions}

In this study, we have investigated the Casimir interaction between ultrathin SWCN films in the dilute regime reporting on a system where materials properties, dimensionality, and temperature have unexpected consequences. SWCN films are inherently anisotropic: when immersed in dielectric layers the quasi-one dimensionality of individual nanotubes asserts the dominance of the response along their lengths. It is thus expected that the Casimir interaction is strongly dependent on the relative optical axis orientation $\varphi$ of two interacting films, as also recently studied in densely packed SWCN films \cite{Bondarev2023}. This giant anisotropy in composite quasi-one dimensional materials then leads us to the notion that Casimir torque in quasi-2D materials is also possible. 

We find that, indeed, the Casimir energy depends on $\varphi$, which drives the emergence of Casimir torque. The interplay between the optical anisotropy and temperature leads to a peculiar separation of quantum mechanical and thermal contributions in the energy and torque. It turns out that thermal fluctuations are especially strong dominating $E$ at sub-$\mu$m separations. The main reason is the reduced dimensionality of the system, which also shows that the particular optical properties (especially the interband terms) play a secondary role in the Casimir interaction. This is consistent with previous studies, which have shown that the reduced dimensionality elevates the importance of thermal fluctuations at smaller separations making the properties of the materials much less important \cite{Drosdoff2014,Le2022}. This is unlike the case of double wall CNTs with inter-tube separations $\sim 3 - 4$ \AA, where the interaction is quantum mechanical and controlled by the specific structure of the nanotube intra and interband optical response contributions \cite{Popescu2011}. 

While thermal fluctuations determine the energy, the Casimir torque, on the other hand, is mostly a quantum mechanical phenomenon. The main reason is attributed to the disappearance of the special $n=0$ Matsubara term in $\mathcal{T}$, a consequence of the quasi-1D anisotropy of the system. We find that the torque in the sub-$\mu$ range is controlled primarily by the intraband contributions in the SWNT optical response and it is exponentially screened by the temperature.

Our results show that in the dilute limit the anisotropic Drude response arising from the quasi-1D SWCN dimensionality is the main reason for the $\varphi$ dependence in the Casimir energy resulting in a relatively strong torque. Qualitatively similar results can be found for other nanotube fims with metallic chiralities. This quantum vs. thermal separation in probing fluctuation induced interactions is a unique feature in metallic nanotube films. We suggest that this peculiar delineation can be studied experimentally, as measurements of the Casimir energy and torque are also possible. For example, our calculations show that, at $T=10$ K, $D = 50$ nm and $\varphi=\frac{\pi}{8}$, $|E|\sim 10.89$ $nJ \cdot m^{-2}$ and $|\mathcal{T}|\sim 8.49$ $nN.m \cdot m^{-2}$, while at $T=300$ K we have $E\sim 9.19$ $nJ\cdot m^{-2}$ and $\mathcal{T}\sim 7.57$ $nN.m\cdot m^{-2}$, which is achievable in the laboratory \cite{Somers2018,Liu2021}. At smaller separations the magnitudes of the Casimir energy and torque are expected to increase due to their scaling laws discussed earlier, which may also be beneficial for potential experimentaion. Our study further shows that investigations of other anisotropic systems at the nanoscale are needed to further understand the interplay between dimensionality, temperature, and materials properties in Casimir phenomena.

\begin{acknowledgments}
We acknowledge support from the US Department of Energy under Grant No. DE-FG02-06ER46297. P. R.-L. acknowledges support from AYUDA PUENTE 2023, URJC and the hospitality of the Theory of Light-Matter and Quantum Phenomena group at the Laboratoire Charles Coulomb, University of Montpellier, where part of this work was done. I.V.B. is supported by the U.S. Army Research Office under award No. W911NF2310206. I.V.B., L.M.W. and M.A. gratefully acknowledge support from the Kavli Institute for Theoretical Physics (KITP), UC Santa Barbara, under U.S. National Science Foundation Grant No. PHY-1748958, where this collaborative work was started. I.V.B. acknowledges KITP hospitality during his invited visit as a KITP Fellow 2022–23 made possible by the Heising-Simons Foundation. M.A. acknowledges the grant ”CAT”, No. A-HKUST604/20, from the ANR/RGC Joint Research Scheme sponsored by the French National Research Agency (ANR) and the Research Grants Council (RGC) of the Hong Kong Special Administrative Region. 
\end{acknowledgments}

\end{document}